\documentclass[journal,twoside,web]{ieeecolor}
\usepackage[table, dvipsnames]{xcolor}
\usepackage{colortbl}

\usepackage{tmi}
\usepackage{cite}
\usepackage{amsmath,amssymb,amsfonts}
\usepackage{algorithmic}
\usepackage{textcomp}

\usepackage[T1]{fontenc}

\usepackage{amsmath}
\usepackage{graphicx}

\usepackage{amssymb}
\usepackage{booktabs}
\usepackage{multirow}
\usepackage{setspace}

\usepackage{orcidlink}
\usepackage{cleveref}

\usepackage[acronym]{glossaries}
\loadglsentries{glossary.sty}

\newcommand{\furl}[1]{{\footnotesize\url{#1}}}

\def\BibTeX{{\rm B\kern-.05em{\sc i\kern-.025em b}\kern-.08em
    T\kern-.1667em\lower.7ex\hbox{E}\kern-.125emX}}
\begin{document}
\title{Temporal Conditioning for Longitudinal Brain MRI Registration and Aging Analysis}
\author{
Bailiang Jian\orcidlink{0009-0009-2419-0420}, 
Jiazhen Pan\orcidlink{0000-0002-6305-8117}, 
Yitong Li\orcidlink{0009-0009-3874-6055}, 
Fabian Bongratz\orcidlink{0009-0009-8879-1823}, 
Ruochen Li\orcidlink{0009-0007-1209-9326},\\
Daniel Rueckert\orcidlink{0000-0002-5683-5889}, 
Benedikt Wiestler\orcidlink{0000-0002-2963-7772}$^*$ 
and Christian Wachinger\orcidlink{0000-0002-3652-1874}$^*$ 
%
\thanks{B. Jian, J. Pan, Y. Li, F. Bongratz, R. Li, D. Rueckert, B. Wiestler, and C. Wachinger are with Technical University of Munich, TUM University Hospital.}%
\thanks{B. Jian, J.Pan, Y. Li, F. Bongratz, D. Rueckert, B. Wiestler, and C. Wachinger are also with Munich Center for Machine Learning (MCML).}%
\thanks{D. Rueckert, B. Wiestler, and C. Wachinger are also with Munich Data Science Institute (MDSI).}%
\thanks{D. Rueckert is also with Imperial College London.}%
\thanks{This work was supported by BMWi (project ``NeuroTEMP'') research funding and MCML. J. Pan is funded by the ERC project Deep4MI (884622).}
\thanks{email: \{bailiang.jian, jiazhen.pan, yi\_tong.li, fabi.bongratz, ruochen.li, daniel.rueckert, b.wiestler, christian.wachinger\}@tum.de}
\thanks{$^*$ B. Wiestler and C. Wachinger contributed equally as senior supervisors.}%
}

\maketitle
\begin{abstract}
Longitudinal brain analysis is essential for understanding healthy aging and identifying pathological deviations. Longitudinal registration of sequential brain MRI underpins such analyses. However, existing methods are limited by reliance on densely sampled time series, a trade-off between accuracy and temporal smoothness, and an inability to prospectively forecast future brain states.
To overcome these challenges, we introduce \emph{TimeFlow}, a learning-based framework for longitudinal brain MRI registration. TimeFlow uses a U-Net backbone with temporal conditioning to model neuroanatomy as a continuous function of age.
Given only two scans from an individual, TimeFlow estimates accurate and temporally coherent deformation fields, enabling non-linear extrapolation to predict future brain states. This is achieved by our proposed inter-/extrapolation consistency constraints applied to both the deformation fields and deformed images. Remarkably, these constraints preserve temporal consistency and continuity without requiring explicit smoothness regularizers or densely sampled sequential data.
Extensive experiments demonstrate that TimeFlow outperforms state-of-the-art methods in terms of both future timepoint forecasting and registration accuracy. 
Moreover, TimeFlow supports novel biological brain aging analyses by differentiating neurodegenerative trajectories from normal aging without requiring segmentation, thereby eliminating the need for labor-intensive annotations and mitigating segmentation inconsistency.
TimeFlow offers an accurate, data-efficient, and annotation-free framework for longitudinal analysis of brain aging and chronic diseases, capable of forecasting brain changes beyond the observed study period. 
\end{abstract}
\section{Introduction}\label{sec:intro}
Understanding how the human brain evolves throughout the lifespan, whether during healthy aging or in the presence of neuropathological disorders such as neurodegenerative diseases, remains a central challenge in neurology and neuroradiology~\cite{nestor2008ventricularad,ott2010adventri}.
Distinguishing these trajectories not only clarifies pathological mechanisms but also refines prognostic models, ultimately supporting timely and personalized interventions~\cite{breijyeh2020adreview}.
Magnetic Resonance Imaging (MRI) has become the standard tool for capturing longitudinal brain changes, owing to its high spatial resolution and superior soft-tissue contrast. 
Repeated imaging sessions allow clinicians and researchers to investigate evolving patterns of atrophy, lesion progression, and other structural alterations over time~\cite{jack2008adni,baheti2021bratsreg}.
\begin{figure}[t]
    \centering
    \includegraphics[width=0.9\linewidth]{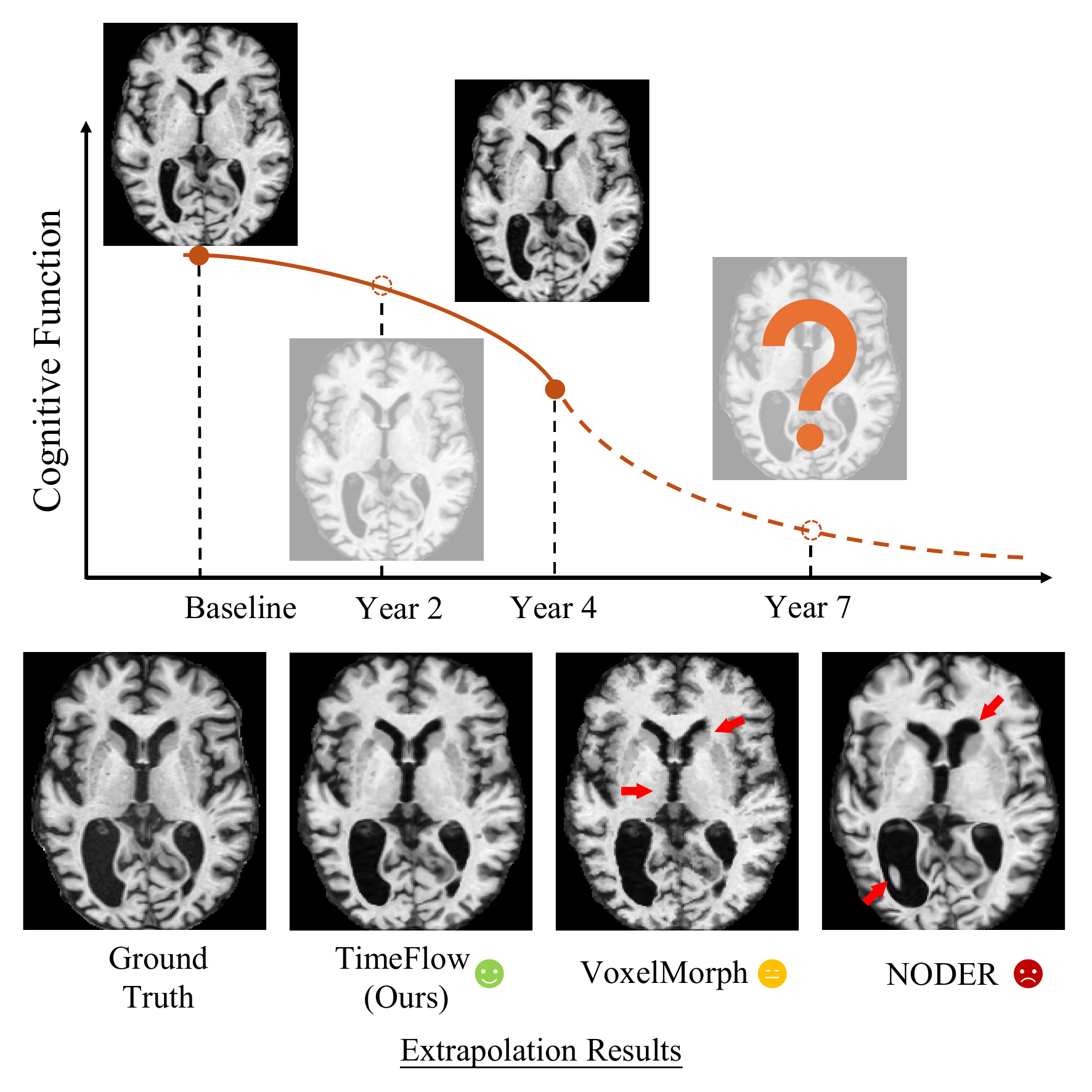}
    \caption{
    \textbf{Top}: Given two longitudinal scans (visits at Baseline and Year 4 visits), TimeFlow can interpolate and extrapolate novel brain MRI scans at unseen timepoints (e.g., Year 2 and Year 7). \textbf{Bottom}: Visual comparison of extrapolated predictions at Year 7 demonstrates TimeFlow's superior performance, compared to (i) NODER, which uses NeuralODE-based geodesic regression, and (ii) VoxelMorph, which linearly scales displacement field. Prediction errors are marked by red arrows.
    }
    \label{fig:intro}
\end{figure}

At the core of most quantitative longitudinal brain analyses lies longitudinal registration, which aligns a subject's brain scans across multiple timepoints~\cite{holland2011nonlinear,simpson2011longitudinal}. Despite steady advances in longitudinal registration, two critical challenges persist:
\textbf{1. Ensuring temporal continuity with limited sequential data.} Existing methods often rely on explicit temporal smoothness constraints~\cite{wu2012temporalfiber,yigitsoy2011temporal,csapo2013temporalsim} which usually compromise accuracy. Besides, these methods often require densely sampled longitudinal data (often more than five scans)~\cite{metz2011nonrigid,qiu2009timelddmm}, which is impractical in typical clinical scenarios where usually fewer than three follow-up scans are available.
\textbf{2. Predicting future brain states and analyzing aging progression.} While retrospective assessments of past changes are informative, prospective forecasting of brain anatomy would markedly enhance clinical utility~\cite{marinescu2019tadpole}. Such forecasting capabilities enable clinicians to anticipate accelerated disease progression and adjust interventions proactively~\cite{lachinov2023diseaseode}. However, most recent methods model only the observed scan sequence~\cite{wu2024seqvxm,bai2024noder,wu2024nodeo-seq}, allowing interpolation between existing timepoints. It limits their capability in extrapolating beyond existing scans, as illustrated in~\Cref{fig:intro}, thereby restricting their ability to assess deviations from expected progression, anticipate disease worsening, and adjust treatment accordingly.

\begin{figure}
    \centering
    \includegraphics[width=\linewidth]{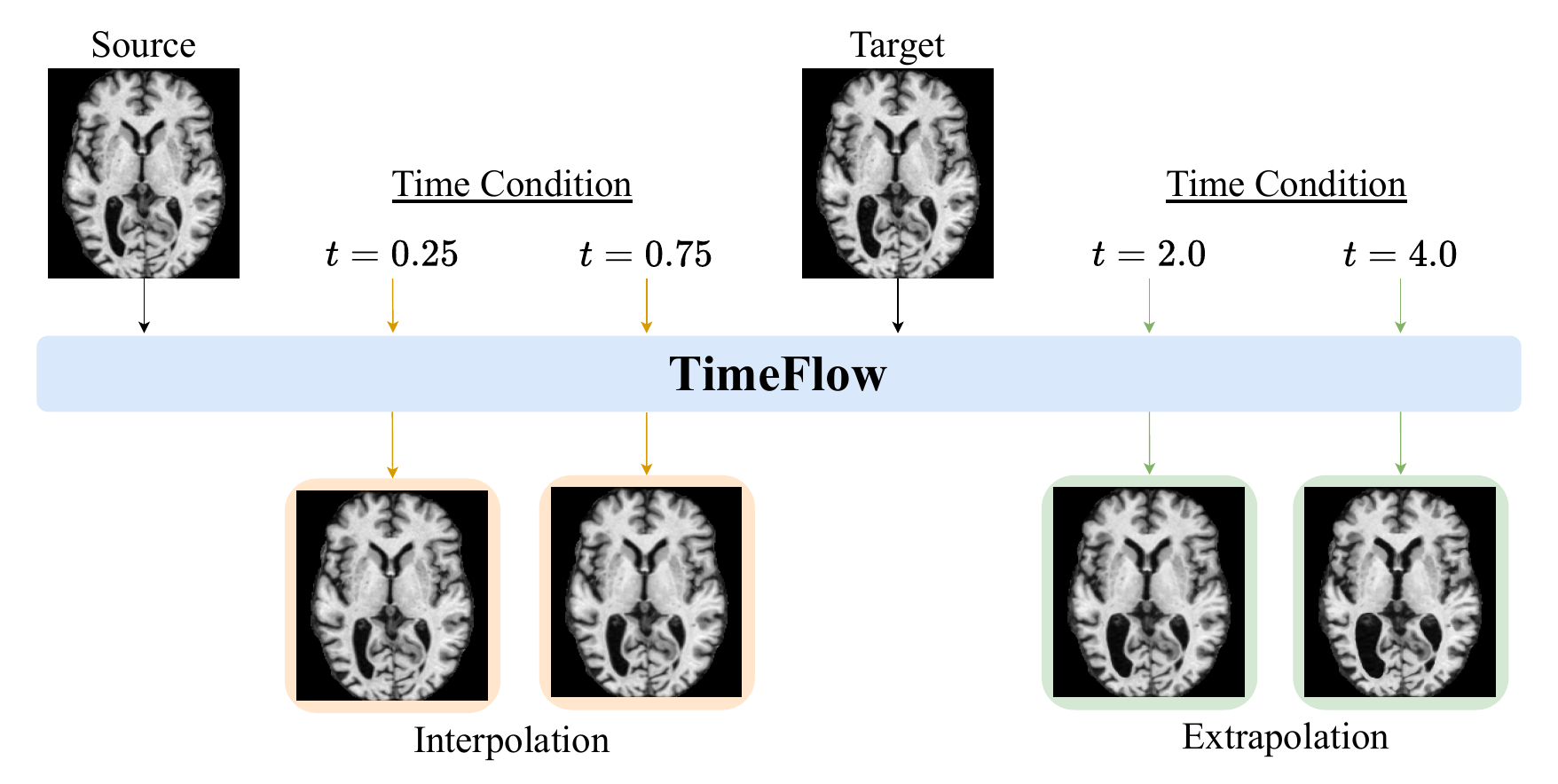}
    \caption{TimeFlow takes source and target images as input, conditioned on the continuous temporal parameter $t$. Adjusting $t$ enables the generation of \textit{interpolated} brain scans within the observed interval ($0 < t < 1$), or \textit{extrapolated} scans for future ($t>1$).}
    \label{fig:taxis}
\end{figure}

To overcome these limitations, we introduce \textit{TimeFlow}, a novel learning-based longitudinal brain MRI registration framework. TimeFlow combines a U-Net backbone with temporal conditioning to model neuroanatomy as a continuous function of age. As shown in~\Cref{fig:taxis}, TimeFlow requires only two scans per subject to estimate the bidirectional deformation fields for arbitrary timepoints, facilitating both interpolation (timepoints within observed intervals) and extrapolation (timepoints beyond observed intervals). Our proposed bidirectional inter-/extra-polation consistency constraints allow unlimited sampling of timepoints, rather than the limited discrete visits during training, ensuring temporally coherent deformation fields. This synergy between the temporal conditioning and the proposed constraints enables TimeFlow to predict future brain states and supports aging analyses without additional annotations. Unlike generative models that create synthetic MRI scans, TimeFlow uses image warping to ensure anatomically plausible outputs, maintain consistency, and provide interpretability through deformation field inspection.

Our contributions are threefold:
\begin{enumerate}
\item \textbf{Temporally-conditioned registration.} A unified network conditioned on a continuous time parameter produces temporally coherent deformation fields from only two input scans, accurately predicting unobserved anatomical states.  
\item \textbf{Bidirectional inter-/extra-polation consistency loss.} Symmetric supervision on bidirectional deformations removes the need for explicit temporal smoothness regularizers, significantly improving temporal coherence, continuity, and generalization through unlimitedly sampled timepoints.
\item \textbf{Registration-only aging analysis.} A registration-only pipeline quantifies relative aging trajectories by referencing healthy controls, differentiating Mild Cognitive Impairment (MCI) and Dementia without requiring tissue labels, thus removing the need for non-trivial annotations.  
\end{enumerate}

\section{Related Work}\label{sec:related_works}
\subsection{Registration across subjects} Cross-sectional registration methods laid the groundwork for medical image registration, pioneering this field with techniques such as parametric spline-based models~\cite{rueckert1999bspline,modat2010fast}, optical flow~\cite{pock2007duality,yang2008fast}, and diffeomorphic transformations~\cite{thirion1998image,vercauteren2009diffeomorphic,beg2005lddmm}. However, they demand prohibitive computing time in the order of several hours to register one single subject, posing a significant barrier to clinical deployment. To address this, learning-based frameworks~\cite{balakrishnan2019voxelmorph,mok2020lapirn,qiu2021vxmbspline,chen2022transmorph,ghahremani2024hvit,jian2024rethinkreg} have emerged, dramatically reducing inference time and often improving registration accuracy. 
Nevertheless, all such methods rely on carefully tuned smoothness regularizers to ensure anatomically plausible deformation fields. An improper weighting factor of smoothness regularizers can either reduce fidelity or introduce unrealistic warps. Additional tuning or adaptive strategies are often needed to strike an optimal balance~\cite{hoopes2021hypermorph,mok2021conditional,wang2023conditional}.

\subsection{Registration across time} Longitudinal registration methods~\cite{shen20044dhammer,hua2008dbm,hua2016dbm,dong2021deepatrophy,dong2024regionaldeepatrophy,qiu2009timelddmm,csapo2013temporalsim,lee2023seq2morph} often leverage the backbone architecture of the cross-sectional registration, whereas one additional constraint is introduced: the temporal continuity/consistency across multiple timepoints of a subject. Two major strategies are often used to tackle this problem. First, an explicit smoothness constraint along the temporal dimension is incorporated to maintain coherent deformation fields over time~\cite{yigitsoy2011temporal,metz2011nonrigid}. However, as with spatial smoothness, temporal regularization faces a similar trade-off: enforcing temporal continuity can inadvertently lower spatial accuracy. Second, many longitudinal methods require multiple, densely sampled scans (>10) to capture temporal dynamics~\cite{bai2024noder,wu2024seqvxm}. In real-world clinical scenarios, patients often have only a few follow-up scans ($\leq 3$), restricting model flexibility and practical implementation. Importantly, these methods commonly can only carry out retrospective analyses and are not capable of prospective studies, such as future image extrapolation and aging/disease progression risk prediction.

\subsection{Longitudinal brain analysis}
Existing longitudinal studies of neurodegenerative diseases like Alzheimer's disease usually require additional annotations, such as cortical surfaces and structural segmentations to quantify the brain changes. However, these labels are usually created by cross-sectional pipelines that process each visit independently, which are both time-consuming and prone to temporal inconsistency, i.e., the same anatomy receives different labels at successive visits. Consequently, quantities derived from those labels (e.g., hippocampal volume, cortical-thickness trajectories) require post-processing to suppress the systematic noise~\cite{leung2010hippovolume,tustison2019corticalthickness}. These limitations motivate annotation-free, temporally coherent models that treat the sequential series as a single spatiotemporal object instead of a collection of disconnected timepoints.
\begin{figure*}[t]
    \centering
    \includegraphics[width=\linewidth]{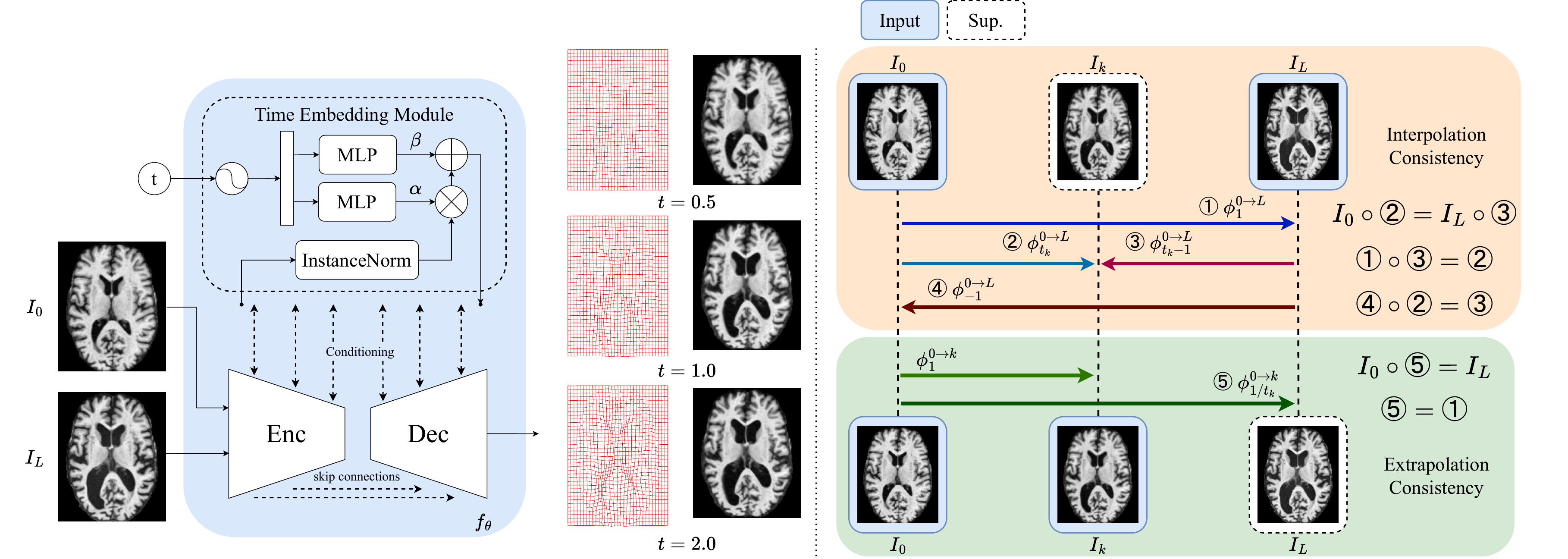}
    \caption{\textbf{Left}: Architecture of \textit{TimeFlow}. The network $f_\theta$ incorporates temporal conditioning via adaptive instance normalization. It takes two input images ($I_0$,$I_L$) and a continuous time variable $t$ to predict the temporal deformation field $\phi_{t}^{0 \rightarrow L}$. Examples of $t=0.5$ (interpolation at middle), $t=1$ (endpoint registration), and $t=2$ (extrapolation) are shown.
    \textbf{Right}: \textit{Interpolation} (upper yellow) and \textit{Extrapolation} (bottom green) Consistency Constraints. The symmetric intermediate image similarity enforces the deformed source and target to be aligned at time $t_k$, while the interpolation flow consistency loss encourages the deformation composition consistency. The extrapolation similarity and flow constraints give information on what a realistic and reasonable forecast should be. Training with both consistency losses enables TimeFlow to generate plausible deformation in $t \in [-\infty, +\infty]$. The input images are highlighted by blue squares (Input), while the supervision is highlighted by dashed squares (Sup.).}
    \label{fig:method}
    \vspace{-0.1cm}
\end{figure*}
\subsection{Future prediction via generative models}
Recent advances in diffusion-based generative models have inspired several attempts to synthesize longitudinal brain MRIs conditioned on variables such as age and diagnosis~\cite{yoon2023sadm,litrico2024tadm}. Despite producing visually plausible synthetic images, these methods largely remain black-box, as their diffusion trajectory does not map clearly onto biophysical deformations. This limits the interpretability, particularly for clinicians interested in tracking local changes like region-specific atrophy. Besides, diffusion models typically require large-scale datasets for stable training, which is impractical in clinical settings where follow-up is sparse. Nonetheless, generative models retain promise as complementary tools in neuropathological disorders involving intensity anomalies, such as gliomas or lesions, where longitudinal registration alone is insufficient to represent newly emerging tissue structures. 

\section{Methods}\label{sec:method}
\subsection{Time-Continuous Registration}
Given a longitudinal MRI sequence $\{I_k\mid k=0, \ldots, L\}$, our goal is to derive a temporally continuous deformation trajectory connecting all intermediate timepoints. To achieve this, we propose a time-continuous registration framework, \emph{TimeFlow}, defined as $\phi_t = f_\theta(I_0, I_L, t)$. As illustrated in~\Cref{fig:taxis}, TimeFlow accepts the baseline image $I_0$ (source) and the last image $I_L$ (target) and predicts the deformation fields $\phi_t$ parametrized by continuous time $t$. 
For interpolation within the observed interval ($0<t<1$), TimeFlow predicts intermediate states between source and target scans. For extrapolation ($t>1$), TimeFlow predicts future brain states beyond the last observed visit.
To enforce temporal coherence and continuity in the deformation trajectories, we introduce the novel inter-/extra-polation consistency constraints. The core architecture (\Cref{fig:method}, Left) is a time-embedded U-Net~\cite{ho2020denoising,rombach2022high}. Specifically, a sinusoidal positional encoding and multi-layer perceptrons (MLPs) project the scalar time variable $t$ into a latent embedding space.
Next, the time latent vector is fed to two separate MLPs to steer the mean and variance of the inter-layer adaptive instance normalization. The network is thus instilled with the capability of generating meaningful time embedding and predicting time-continuous flow fields by the novel time conditioning mechanism. 

\subsection{Interpolation Consistency}
\paragraph{\textbf{Symmetric Intermediate Similarity}}
We first define the time-continuous registration function as $\phi_t^{0 \rightarrow L} = f_\theta(I_0, I_L, t)$, where $f_\theta(I_0, I_L, 0) = \textrm{Id}$ is the identity mapping at $t=0$.
When $t \in (0,1]$, $\phi_t^{0 \rightarrow L}$ is the forward deformation mapping from $I_0$ towards $I_L$, and $\phi_1^{0 \rightarrow L}$ moves $I_0$ to reach $I_L$. Similarly, for negative values, $t<0$,  $\phi_t$ maps backward from $I_L$ to $I_0$.
For any intermediate image $I_k$ observed at time $t_k$, we can formulate the \textit{symmetric interpolation image similarity} measure as: 
\begin{equation}\label{eq:sim_mid}
\mathcal{L}_\textrm{sim}(I_0 \circ \phi_{t_k}^{0 \rightarrow L}, I_k) + \mathcal{L}_\textrm{sim}(I_L \circ \phi_{t_k - 1}^{0 \rightarrow L}, I_k),
\end{equation}
which deforms both source and target images to remove the bias introduced by single-side warping~\cite{avants2008syn,mok2020symnet,hoffmann2024synthMorph}. 
As shown in~\Cref{fig:method} (Right, top yellow), $I_0 \circ \phi_{t_k}^{0 \rightarrow L}$ is warping $I_0$ to the time $t=t_k$ and $I_L \circ \phi_{t_k - 1}^{0 \rightarrow L}$ is warping $I_L$ to the time $t=t_k$.
This symmetry enforces bidirectional consistency, facilitating plausible deformation trajectories passing through observed intermediate scans.

\paragraph{\textbf{Generalized Symmetric Intermediate Similarity}}\label{sec:sample_t}
However, in real-world clinical scenarios, many subjects have fewer than three MRI scans available. 
In such a case, we generalize the above constraint by synthetically sampling intermediate timepoints $\hat{t}$ from a uniform distribution $\mathcal{U}(0,1)$, defining the \textit{generalized interpolation similarity} as:
\begin{equation}\label{eq:sim_mid_v2}
    \mathcal{L}_\textrm{sim}(I_0 \circ \phi_{ \hat{t} }^{0 \rightarrow L}, I_L \circ \phi_{ \hat{t} - 1}^{0 \rightarrow L}),    
\end{equation} 
Here, both warped images $I_0 \circ \phi_{ \hat{t} }^{0 \rightarrow L}$ and $I_L \circ \phi_{ \hat{t} - 1}^{0 \rightarrow L}$ serve as a synthetic version of intermediate timepoint $I_{\hat{t}}$, eliminating reliance on actual intermediate observations. 
By freely sampling $\hat{t}$, TimeFlow effectively learns densely and continuously across the entire temporal axis.

\paragraph{\textbf{Interpolation Flow Consistency}}
To further encourage plausible time-continuous deformations, we propose the \textit{interpolation flow consistency} as a regularizer:
\begin{equation}\label{eq:flow_mid}
\begin{split}
    \mathcal{L}_\textrm{inter-consis-flow} = & \|\phi_1^{0 \rightarrow L} \circ \phi_{t_k-1}^{0 \rightarrow L} - \phi_{t_k}^{0 \rightarrow L} \|_2^2 \\ + & \|\phi_{-1}^{0 \rightarrow L} \circ \phi_{t_k}^{0 \rightarrow L} - \phi_{t_k-1}^{0 \rightarrow L}\|_2^2,
\end{split}
\end{equation}
which combines the two constraints shown in~\Cref{fig:method} (Right, top yellow).
We formulated the loss with the notation $t_k$, but as described in the previous paragraph, $t_k$ can also be sampled from the uniform distribution if no intermediate observation exists.
It is derived from the common transformation composition consistency that two consecutive transforms are equivalent to a direct transform. Moving from $t=0$ to $t=1$ and then back to $t=t_k$ is equivalent to moving directly from $t=0$ to $t=t_k$ (arrow 1 composed with arrow 3 should equal to arrow 2 in~\Cref{fig:method}, Right). The backward case is derived similarly. It is worth noting that this differs from the commonly used smoothness regularizations, which explicitly enforce a smooth gradient over the flow field. Therefore, there is only a minimal trade-off between accuracy and smoothness since both the symmetric image similarity and the flow consistency encourage plausible and reasonable longitudinal transformations.

\subsection{Extrapolation Consistency}
Through the intermediate similarity and interpolation flow consistency, we obtain meaningful predictions for bidirectional registrations in $t \in [-1, 1]$.
Now we introduce extrapolation constraints for forward ($t>1$) and backward ($t<-1$) directions.
To this end, we consider the deformation $\phi_1^{0 \rightarrow k}$ from $I_0$ to $I_k$. 
Given the time of $I_k$ being $t_k$, it holds that $\phi_{t_k}^{0 \rightarrow L} = \phi_1^{0 \rightarrow k}$. As illustrated in~\Cref{fig:method} (Right), they both register $I_0$ to $I_k$. 
Consequently, we can then formulate $\phi_{1}^{0 \rightarrow L} = \phi_{1/t_k}^{0 \rightarrow k}$ (arrow 1 equals to arrow 5 in~\Cref{fig:method}, Right), where $\phi_{1/t_k}^{0 \rightarrow k}$ denotes an extrapolation because $t_k \in (0,1)$ and $1/t_k>1$.  
Based on this notation, we define the \textit{extrapolation image similarity} as:
\begin{equation}\label{eq:sim_ext}
    \mathcal{L}_\textrm{sim}(I_0 \circ \phi_{1/t_k}^{0 \rightarrow k}, I_L).
\end{equation}
Again, if no intermediate observations are available, we can sample $t_k$ from $\mathcal{U}(0,1)$ and define $I_k = I_0 \circ \phi_{t_k}^{0 \rightarrow L}$ to create synthetic intermediate data. 
In addition to the extrapolation similarity measure, we define the forward \textit{extrapolation flow consistency} as:
\begin{equation}
    \mathcal{L}_\textrm{ext-consis-flow} = \|\phi_1^{0 \rightarrow L} - \phi_{1/t_k}^{0 \rightarrow k} \|_2^2. 
\end{equation}
As shown in~\Cref{fig:method} (Right), both $\phi_1^{0 \rightarrow L}$ (arrow 1) and $\phi_{1/t_k}^{0 \rightarrow k}$ (arrow 5) map from $I_0$ to $I_L$, thus should be consistent with each other.
The backward versions of the \textit{extrapolation similarity} and \textit{extrapolation flow consistency} are similarly defined as:
\begin{equation}
\begin{split}
    &\mathcal{L}_\textrm{sim}(I_L \circ \phi_{1/(t_k-1)}^{k \rightarrow L}, I_0) \quad \text{and} \\
    &\mathcal{L}_\textrm{ext-consis-flow} =\|\phi_{-1}^{0 \rightarrow L} - \phi_{1/(t_k-1)}^{k \rightarrow L}\|_2^2.
\end{split}
\end{equation}
The similarity measures and flow consistencies both contribute to achieving a realistic and coherent extrapolation forecasting. The proposed losses effectively ensure accurate extrapolation while implicitly capturing temporal consistency and continuity, without the need for additional temporal smoothness constraints.

\subsection{Unlimited Sampled Triplets}
As illustrated in~\Cref{fig:method} (Right), the proposed inter-/extra-polation constraints require observed triplets of timepoints, limiting learning when few discrete scans are available. To address this, we use ``synthetic'' triplets dynamically sampled from continuous time intervals. Specifically, by sampling intermediate timepoints from uniform distribution $\hat{t}\sim\mathcal{U}(0,1)$ during training, we can generate infinite triplets from a pair of source and target images, freeing TimeFlow from the discrete, fixed timepoints. This continuous, densely sampled temporal supervision significantly enhances the network's capacity to generalize and accurately model brain anatomy changes at any arbitrary timepoint, including unseen future scenarios, while using only two scans as inputs.

\section{Data and Experiments}\label{sec:exp}

\paragraph{\textbf{Data and Pre-processing}} 
We train and evaluate TimeFlow on the publicly available \gls{ADNI} dataset~\cite{jack2008adni}, which comprises longitudinal T1-weighted \gls{MR} scans from both cognitively normal subjects and patients diagnosed with \gls{AD} or \gls{MCI}, i.e., neurodegenerative disorders characterized by progressive patterns of brain atrophy. 
All scans undergo standardized pre-processing using the longitudinal pipeline in FreeSurfer v7.2~\cite{fischl2012freesurfer}. Each scan is first processed individually for skull stripping, normalization. Subsequently, all scans from the same subject are rigidly aligned via a groupwise rigid registration~\cite{reuter2012freesurfertemplate}. FreeSurfer aggregates information across timepoints and refines the pre-processed outputs. 
To ensure meaningful longitudinal analysis, we select 134 subjects exhibiting substantial structural changes (median brain deformation $>0.3$mm).
The dataset is partitioned into training (74 subjects), validation (30 subjects), and testing (30 subjects) sets at the subject level prior to any processing, ensuring complete subject disjointness across splits and preventing information leakage. The splits are constructed to maintain balance across diagnosis, age, sex, and number of available visits. Each subject contributes to 2 to 12 longitudinal scans. All images are resampled to isotropic 1mm spacing and cropped to a uniform size of $160\times160\times192$ voxels. Intensity values are normalized to the range [0,1] based on the [0, 99.9] percentile without clipping.
Additionally, 11 subjects from the OASIS dataset~\cite{lamontagne2019oasis} are used for zero-shot evaluation to assess TimeFlow's generalization to data from different scanners and imaging protocols.

\begin{figure}[t]
    \centering
    \includegraphics[width=\linewidth]{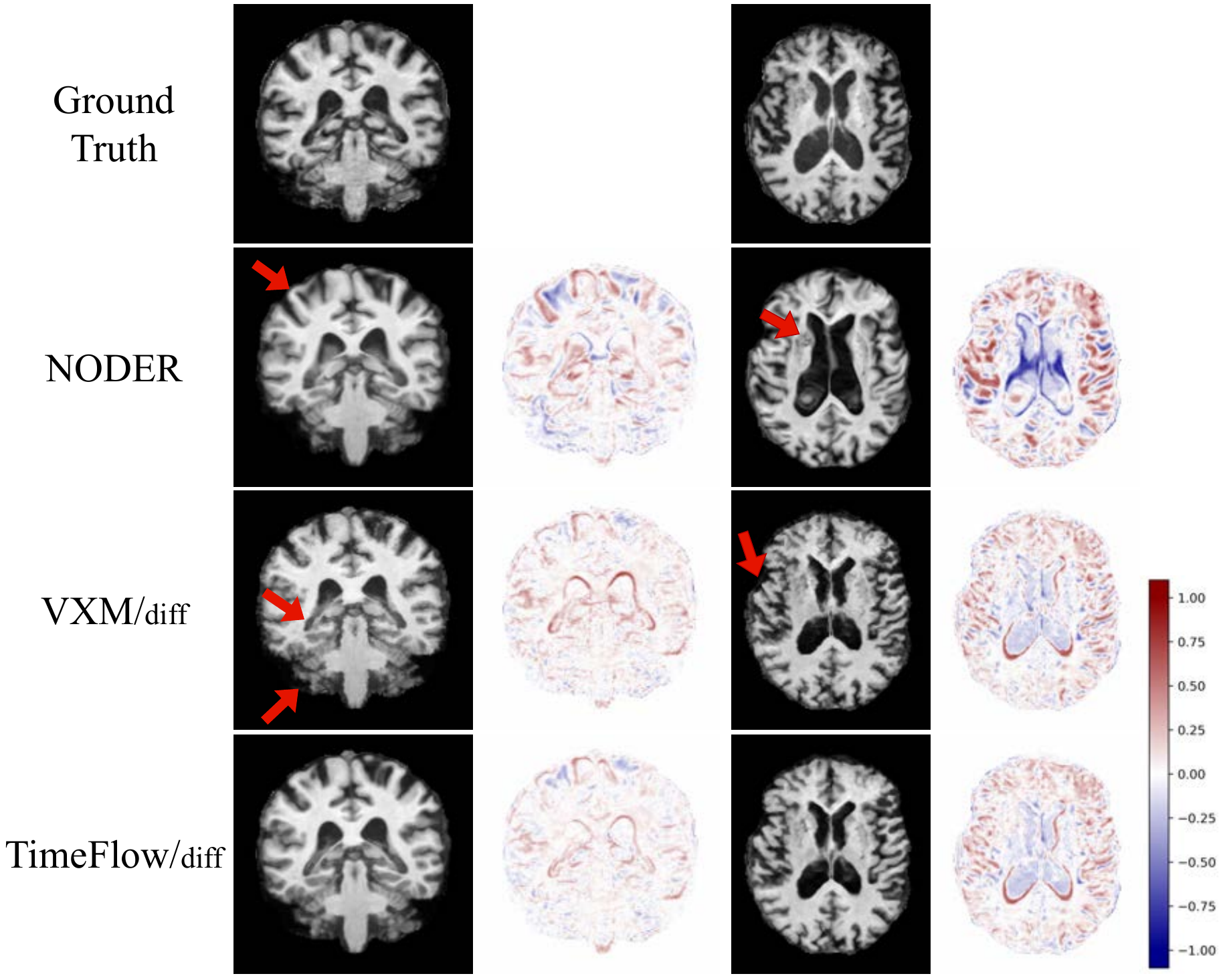}
    \caption{Visualization of extrapolation errors compared to ground truth future images. Coronal and axial views of two subjects are presented. Areas with high error are marked by red arrows.}
    \label{fig:qual_ext_left}
    \vspace{-0.3cm}
\end{figure}

\paragraph{\textbf{Implementation Details}} Our model is implemented using Pytorch 2.1.2. and training/test is carried out on one NVIDIA A100 40G with a batch size of $2$ (sample $\hat{t}$ twice per image pair). We set the dimension of time embedding to $16$, the feature channels in the U-Net encoder and decoder to $[32,32,48,48,96]$. In addition to TimeFlow, which outputs the time-dependent deformation field directly, we have the diffeomorphic variant TimeFlow\textsubscript{diff}, which utilizes the \textit{scaling and squaring} layer with $7$ integration steps to integrate the temporal stationary velocity field. The weights of both interpolation and extrapolation similarity are set to $1.0$. The weights of interpolation and extrapolation flow consistency are set to $2, 0.03$ for TimeFlow and $1.25, 0.025$ for TimeFlow\textsubscript{diff}.

\begin{figure}[t]
    \centering
    \includegraphics[width=\linewidth]{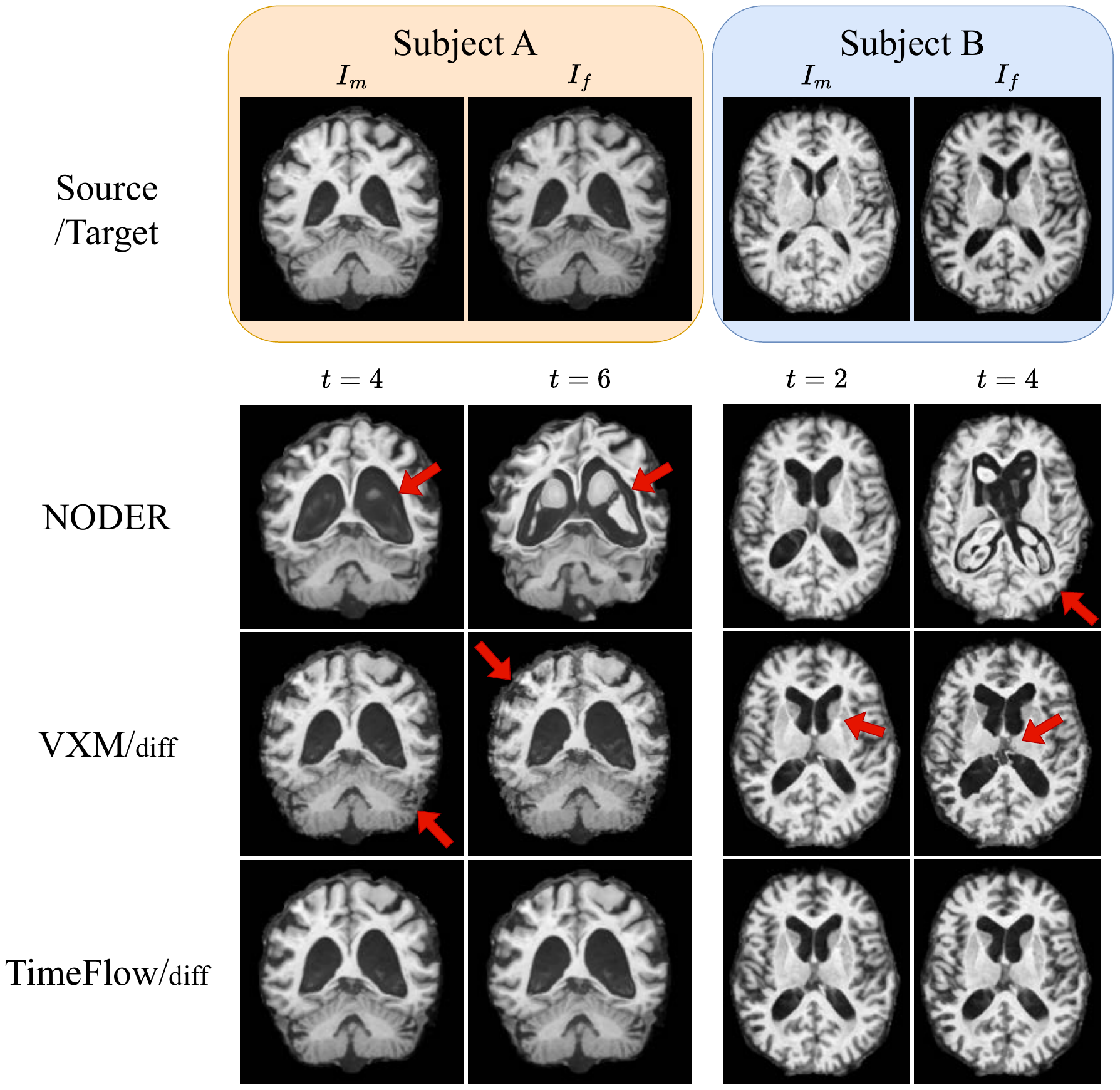}
    \caption{Future image predictions at different extrapolation time $t$, where no ground truth is available. $t=1$ represents one-step pairwise registration between $I_m$ and $I_f$. Thus, $t=2,4,6$ represents two-, four-, and six-step extrapolation over $I_f$. Major errors are highlighted with red arrows. VXM/\textsubscript{diff} produces unsmooth deformations, leading to structural inconsistencies and loss of anatomical details (zoom in for more details).}
    \label{fig:qual_ext_right}
    \vspace{-0.3cm}
\end{figure}

\begin{table*}[ht]
\centering
\setlength{\tabcolsep}{5pt}
\caption{Comparative evaluation of \textit{TimeFlow/\textsubscript{diff}} versus baseline methods in terms of \gls{MAE} (in scale $\times 10^{-2}$), \gls{PSNR}, SD$\log$J (in scale $\times 10^{-2}$) and non-diffeomorphic volume (NDV) in the foreground brain area. \textit{Seq-VXM}, which performs multiple pairwise registrations, represents an upper bound for interpolation performance. Methods with the subscript \textsubscript{diff} use the \textit{scaling and squaring} method to integrate the \gls{SVF}. Best scores are marked in \textbf{bold}.}
\begin{tabular}{llllllllll}
    \toprule
    & & \multicolumn{4}{c}{Extrapolation} & \multicolumn{4}{c}{Interpolation} \\
    \cmidrule(lr){3-6} \cmidrule(lr){7-10}
    Dataset & Method & MAE $\downarrow$ & PSNR $\uparrow$ & SD$\log$J $\downarrow$ & NDV $\downarrow$ 
            & MAE $\downarrow$ & PSNR $\uparrow$ & SD$\log$J $\downarrow$ & NDV $\downarrow$ \\
    \midrule
    \multirow{9}{*}{ADNI}
    & Seq-VXM & \multicolumn{4}{c}{\multirow{2}{*}{-}} & \multirow{2}{*}{4.5$\pm$0.9} & \multirow{2}{*}{25.2$\pm$1.7} & \multirow{2}{*}{2.6$\pm$0.3} & \multirow{2}{*}{0.0$\pm$0.1} \\
    & (Upper bound) & & & & & & & & \\
    \addlinespace[0.1em]
    \cmidrule(lr){2-10}
    & rigid & 10.3$\pm$1.9 & 16.9$\pm$1.5 & - & - & 7.3$\pm$1.1 & 20.6$\pm$1.5 & - & - \\
    \addlinespace[0.1em]
    & SyN & 9.3$\pm$1.9  &  18.1$\pm$1.8  &  4.5$\pm$2.6  &  $10^2 \pm 10^3$  &  5.7$\pm$1.0  &  22.7$\pm$1.6  &  1.3$\pm$0.3  &  0.7$\pm$3.0  \\
    \addlinespace[0.1em]
    & NODER & 9.3$\pm$1.9 & 18.0$\pm$1.7 & \textbf{1.8$\pm$0.9} & \textbf{0.2$\pm$1.9} & 5.5$\pm$0.9 & 23.2$\pm$1.5 & 1.8$\pm$0.5 & 0.0$\pm$0.3 \\
    \addlinespace[0.1em]
    & VXM & 9.1$\pm$1.9 & 18.3$\pm$1.9 & 6.3$\pm$4.8 & $10^4\pm10^4$ & 5.6$\pm$1.1 & 22.9$\pm$1.7 & \textbf{1.5$\pm$0.2} & 0.2$\pm$0.4 \\
    \addlinespace[0.1em]
    & VXM\textsubscript{diff} & 9.2$\pm$2.0 & 18.4$\pm$1.9 & 6.3$\pm$2.0 & $10^3\pm10^3$ & 5.5$\pm$1.0 & 23.2$\pm$1.7 & 2.2$\pm$0.3 & 0.1$\pm$0.2 \\
    \addlinespace[0.1em]
    \cmidrule(lr){2-10}
    & TimeFlow & \textbf{8.6$\pm$1.6} & \textbf{18.8$\pm$1.7} & 5.0$\pm$1.0 & $10^3\pm10^3$ & 5.5$\pm$1.0 & 23.0$\pm$1.7 & 1.6$\pm$0.1 & 0.1$\pm$0.4 \\
    \addlinespace[0.1em]
    & TimeFlow\textsubscript{diff} & 8.8$\pm$1.8 & 18.7$\pm$1.9 & 6.0$\pm$1.0 & 6.1$\pm$11.2 & \textbf{5.3$\pm$1.0} & \textbf{23.3$\pm$1.8} & 2.2$\pm$0.2 & \textbf{0.0$\pm$0.0} \\
    \addlinespace[0.1em]
    \midrule
    \multirow{7}{*}{\shortstack{OASIS\\(Zero-shot)}}
    & rigid & 12.7$\pm$2.0 & 15.1$\pm$0.9 & - & - & 8.9$\pm$1.1 & 18.2$\pm$1.1 & - & - \\
    \addlinespace[0.1em]
    & SyN & 10.9$\pm$2.2  &  16.7$\pm$1.7  &  4.6$\pm$2.2  &  $10^4\pm10^4$  &  7.3$\pm$1.3  &  20.1$\pm$1.6  &  1.7$\pm$0.2  &  0.3$\pm$0.9  \\
    \addlinespace[0.1em]
    & NODER & 11.0$\pm$2.2 & 16.5$\pm$1.4 & \textbf{2.9$\pm$0.9} & \textbf{0.7$\pm$2.9} & \textbf{6.9$\pm$1.4} & \textbf{21.0$\pm$1.6} & 2.4$\pm$0.5 & \textbf{0.0$\pm$0.1} \\
    \addlinespace[0.1em]
    & VXM & 10.7$\pm$2.3 & 16.8$\pm$1.7 & 5.4$\pm$3.8 & $10^4\pm10^4$ & 7.3$\pm$1.3 & 20.3$\pm$1.5 & \textbf{1.7$\pm$0.2} & 0.1$\pm$0.2 \\
    \addlinespace[0.1em]
    & VXM\textsubscript{diff} & 10.4$\pm$2.2 & 17.1$\pm$1.7 & 6.3$\pm$2.0 & $10^3\pm10^3$ & 7.1$\pm$1.3 & 20.5$\pm$1.6 & 2.7$\pm$0.3 & 0.1$\pm$0.4 \\
    \addlinespace[0.1em]
    \cmidrule(lr){2-10}
    & TimeFlow & 10.2$\pm$1.8 & 17.1$\pm$1.5 & 5.1$\pm$1.4 & $10^3\pm10^3$ & 7.2$\pm$1.2 & 20.3$\pm$1.6 & 1.8$\pm$0.2 & 0.1$\pm$0.3 \\
    \addlinespace[0.1em]
    & TimeFlow\textsubscript{diff} & \textbf{9.8$\pm$1.7} & \textbf{17.5$\pm$1.5} & 6.2$\pm$1.1 & 26.8$\pm$56.6 & 7.0$\pm$1.2 & 20.5$\pm$1.6 & 2.6$\pm$0.3 & 0.0$\pm$0.2 \\
    \addlinespace[0.1em]
    \bottomrule
\end{tabular}
\label{tab:reg_results}
\end{table*}

\begin{figure*}[ht!]
    \centering
    \includegraphics[width=\linewidth]{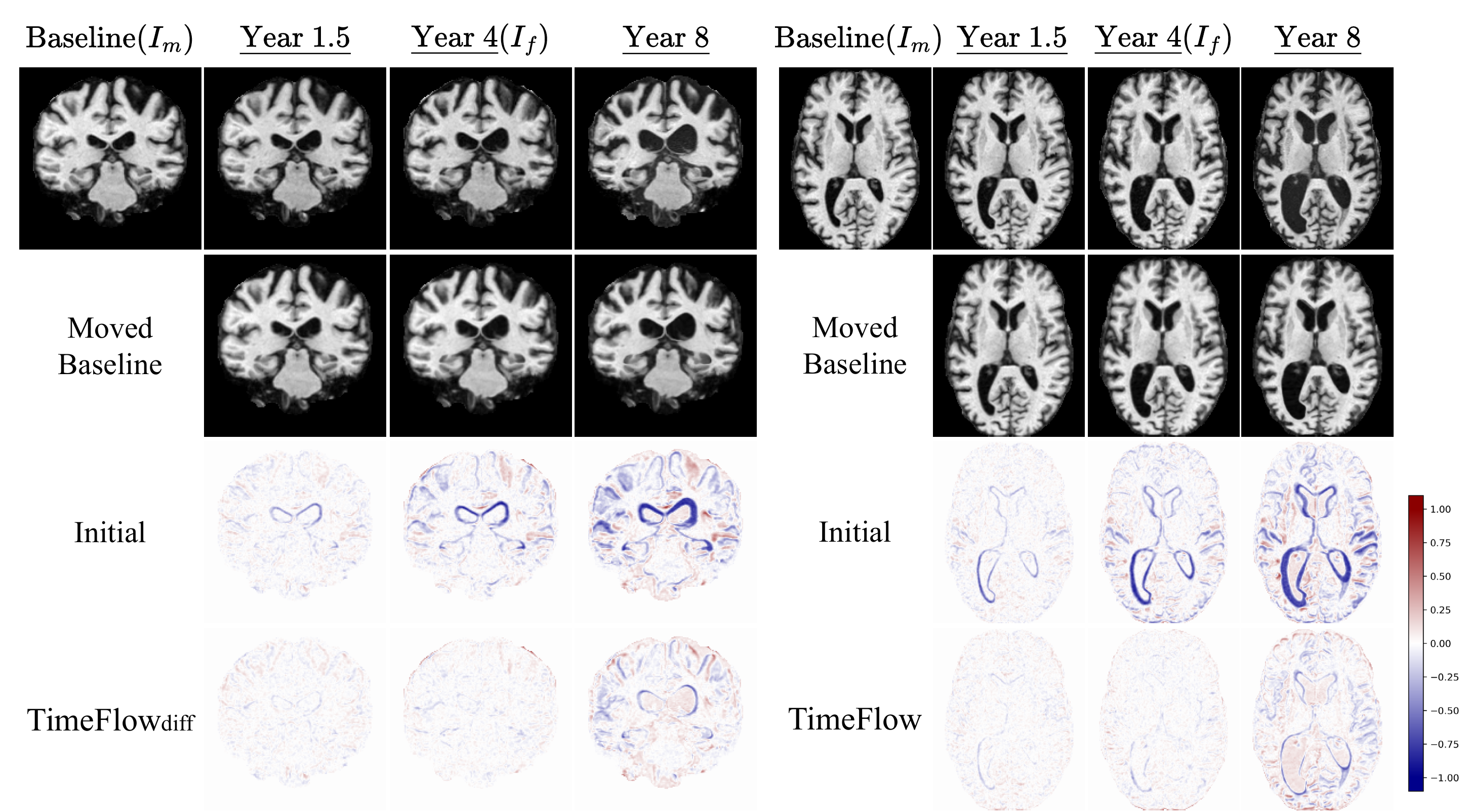}
    \caption{Visualization of interpolation (Year 1.5) and extrapolation (Year 8) results from coronal and axial views of a subject's brain. \textit{TimeFlow/\textsubscript{diff}} takes the \underline{Baseline} and \underline{Year 4} images as input, interpolates to \underline{Year 1.5}, and extrapolates to \underline{Year 8}. Ground truth images (row 1) and warped images (row 2) are compared. Initial misalignment (row 3) is effectively corrected (row 4), demonstrating TimeFlow's accuracy in temporal modeling. Error values range from $[-1, 1]$ due to the image intensity normalization to $[0,1]$.}
    \label{fig:qual}
\end{figure*}

\paragraph{\textbf{Baselines}}\label{sec:baseline} To evaluate TimeFlow's registration performance, we compare against the following methods:

\begin{itemize}
    \item[\textbullet] NODER~\cite{wu2024nodeo-seq,bai2024noder} takes the complete image sequence as input and is thus aware of every timepoint. It performs subject-specific optimization and models the time-varying velocity field with a neural network and leverages NeuralODE~\cite{chen2018neuralode} to compute integration and adjoint gradients. It fits the subject-specific temporal deformation trajectory by maximizing the similarity with each frame.

    \item[\textbullet] VoxelMorph (VXM)~\cite{balakrishnan2019voxelmorph} and the diffeomorphic variant (VXM\textsubscript{diff})~\cite{dalca2019diffvxm}. Since both are pairwise registration methods, the interpolation and extrapolation prediction is carried out based on the linear scaling of the displacement or velocity field from the pairwise registration. The number of integration steps of VXM\textsubscript{diff} is set to 7.

    \item[\textbullet] SyN (ANTs)~\cite{avants2008syn}. We apply the symmetric diffeomorphic image pairwise registration algorithm with normalized cross-correlation (radius $r=3$). For interpolation and extrapolation, displacement fields are linearly scaled, analogous to VXM.

    \item[\textbullet] Sequence-based VoxelMorph (Seq-VXM)~\cite{wu2024seqvxm}. Seq-VXM processes the whole sequence and registers the baseline image to each intermediate timepoint through multiple pairwise registration. Since Seq-VXM uses the full image sequence as input, we treat it as the \emph{performance upper bound} for the retrospective interpolation task (refer to Tab.~\ref{tab:reg_results}).
\end{itemize}

All methods use \gls{LNCC} with radius $r=9$ (NODER) or $r=4$ (others) as image similarity loss.
The loss function weighting factors for NODER follow the original implementation. Seq-VXM, VXM and VXM\textsubscript{diff} compute the SOTA smoothness regularizer GradICON~\cite{tian2023gradicon,tian2024unigradicon} with weight $1.5, 0.5, 1.0$.

\begin{figure*}[t]
    \centering
    \includegraphics[width=0.9\linewidth]{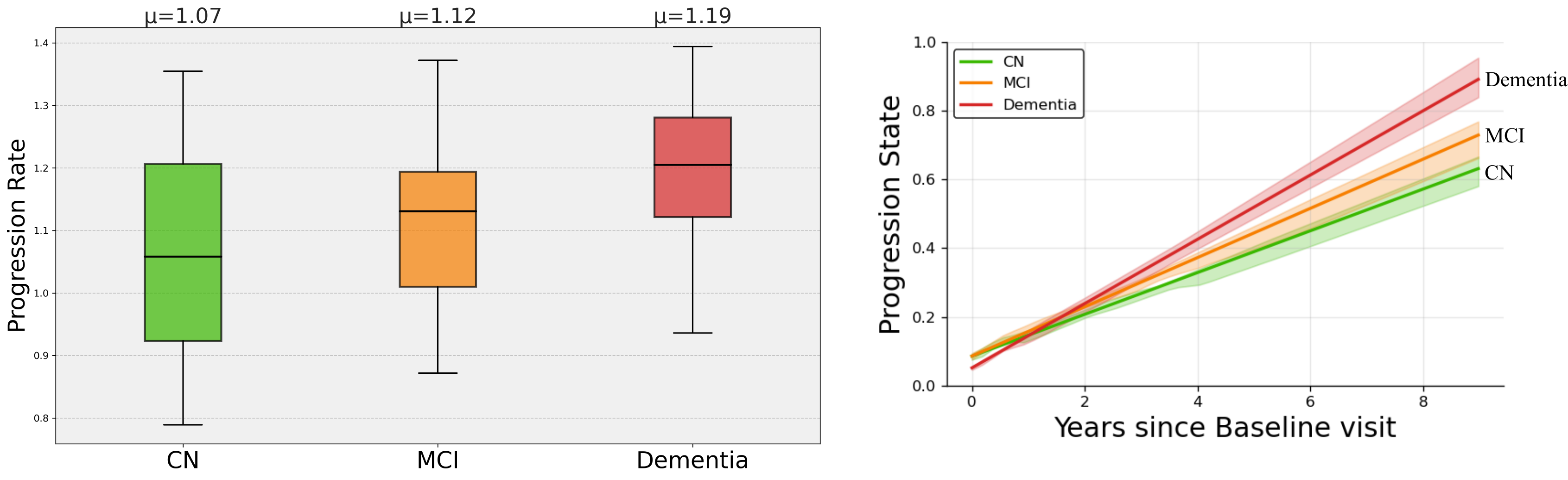}
    \caption{\textbf{Left}: Prospective, annotation-free analysis of brain aging progression rates for Control (CN), \gls{MCI} and Dementia. Rates near 1 reflect normal, chronological brain aging, whereas values above 1 indicate accelerated neurodegenerative progression.
    \textbf{Right}: Retrospective group-wise analysis of longitudinal brain progression. Trajectories represent median linear fits between TimeFlow's inferred temporal brain state ($t$) and chronological years since baseline. Shaded ribbons indicate the 40th-60th percentile range. A steeper slope corresponds to faster neuroanatomical changes. Colormap: CN-green, MCI-orange, Dementia-red.
    }
    \label{fig:progress}
\end{figure*}
\begin{figure}[ht]
    \centering
    \includegraphics[width=\linewidth]{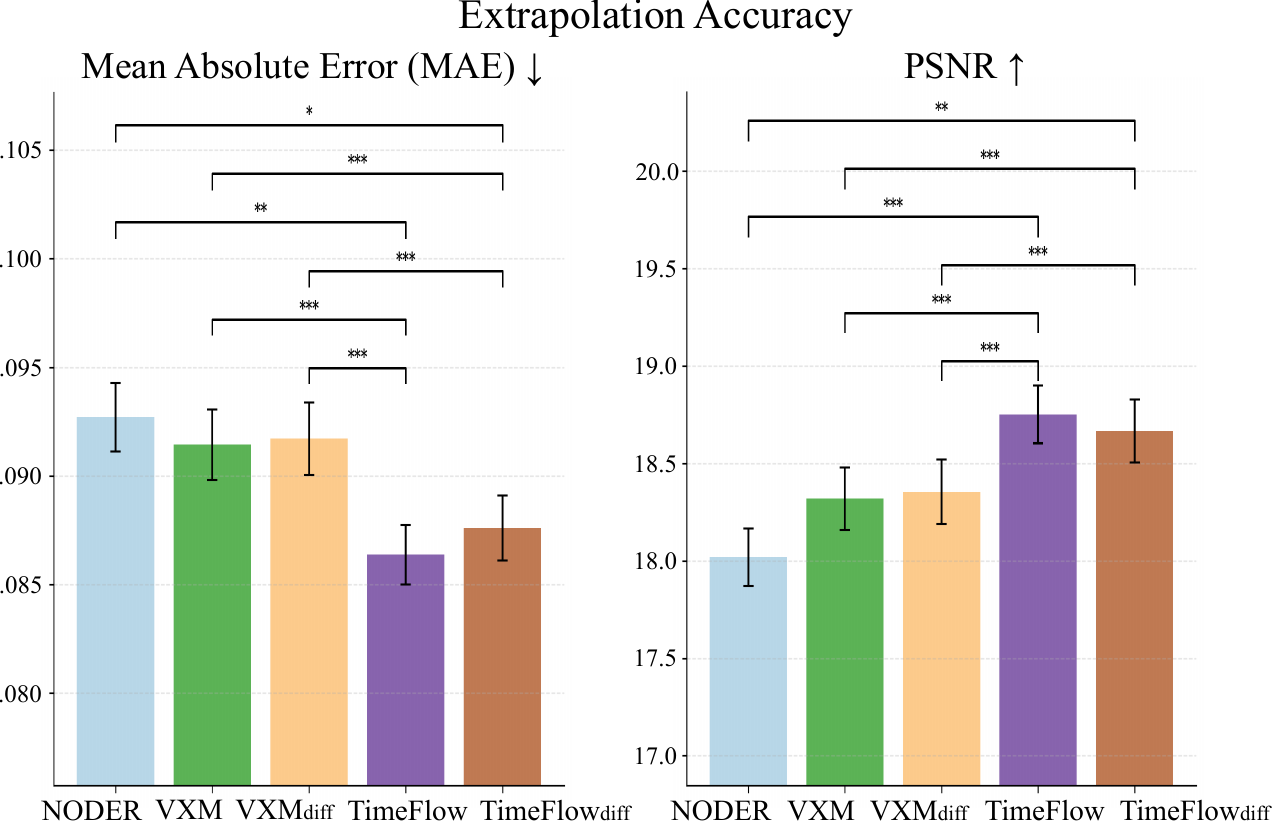}
    \caption{Quantitative comparison of extrapolation accuracy among TimeFlow and baseline methods on ADNI dataset. Barplots represent mean values with standard errors. Significance levels by Wilcoxon $t$-test (Holm–Bonferroni correction) are indicated. ***:p<0.001, **:p<0.01, *:p<0.05}
    \label{fig:ext_bars}
\end{figure}

\begin{table}[t]
\setlength{\tabcolsep}{5pt}
\centering
\caption{Dice Score on ADNI dataset. For each method, we report the mean Dice scores and standard deviation across all subjects and anatomical regions. Higher values indicate better alignment.}
\label{tab:adni_dice}
\begin{tabular}{lcc}
    \toprule
    Method & Extrapolation & Interpolation \\
    \midrule
    \addlinespace[0.1em]
    Seq-VXM & - & 81.17 $\pm$ 1.85 \\
    \midrule
    \addlinespace[0.1em]
    SyN & 72.85 $\pm$ 6.03 & 80.32 $\pm$ 2.76 \\
    \addlinespace[0.1em]
    NODER & 75.43 $\pm$ 4.43 & 80.43 $\pm$ 1.88 \\
    \addlinespace[0.1em]
    VXM & 75.78 $\pm$ 4.48 & 81.15 $\pm$ 3.14 \\
    \addlinespace[0.1em]
    VXM\textsubscript{diff} & 75.35 $\pm$ 4.50 & 81.11 $\pm$ 2.75 \\
    \addlinespace[0.1em]
    TimeFlow & 76.58 $\pm$ 3.90 & \textbf{81.19 $\pm$ 2.88} \\
    \addlinespace[0.1em]
    TimeFlow\textsubscript{diff} & \textbf{76.95 $\pm$ 3.80} & 81.10 $\pm$ 2.65 \\
    \bottomrule
\end{tabular}
\end{table}

\paragraph{\textbf{Evaluation metrics}} We employ both intensity-based and segmentation-based metrics to comprehensively evaluate registration performance. 
For intensity-based evaluation, we use mean absolute error (MAE) and peak signal-to-noise ratio (PSNR) to quantitatively evaluate the image alignment. To assess the plausibility and smoothness of deformation fields, we compute the standard deviation of the logarithm of the Jacobian determinant (SD$\log$J)~\cite{leow2007sdlogj} and non-diffeomorphic volume (NDV)~\cite{liu2024ndv}, which measure irregular volume change and detect non-diffeomorphic deformations. The aforementioned metrics are only applied in the foreground brain area.
For segmentation-based evaluation, we report Dice scores computed on anatomical structures segmented by FreeSurfer.

\section{Results}
\subsection{Prospective studies: Predicting the future}
\begin{figure*}[h]
    \centering
    \includegraphics[width=\linewidth]{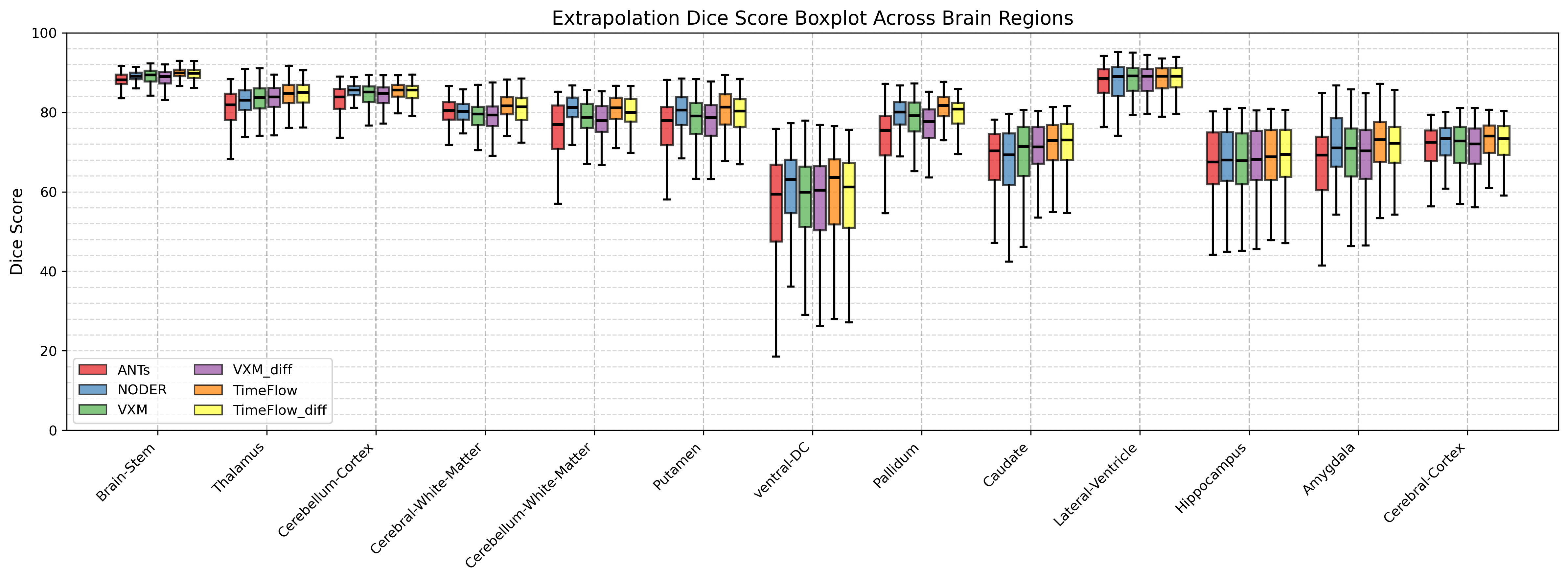}
    \caption{Region-wise Dice score comparison for \textbf{extrapolation} on ADNI dataset. Each boxplot shows the distribution of Dice scores across subjects for major anatomical structures. TimeFlow variants achieve consistently higher or comparable Dice scores across all regions.}
    \label{fig:ext_dice_box}
    \vspace{-0.15cm}
\end{figure*}

We first evaluate TimeFlow's performance in prospective scenarios, specifically, its ability to extrapolate deformation fields to predict future brain scans. TimeFlow supports two forms of prospective analysis: (1) Given two available scans and a future time variable $t>1$, the model directly predicts the corresponding deformation fields and resulting future MRI images; (2) Given an observed future scan, TimeFlow can infer the corresponding $\phi_t$ and use the estimated time variable $t$ as a quantitative metric to quantify the biological aging progression.
\paragraph{\textbf{Given t, predict the future image}} A significant advantage of TimeFlow, as described in~\Cref{sec:method}, is its ability to extrapolate deformation fields to future timepoints beyond the observed interval. The quantitative performance of TimeFlow's extrapolation forecasting is reported in~\Cref{tab:reg_results}. The metrics are computed with respect to the ground truth future scans. Results clearly demonstrate that TimeFlow consistently achieves superior extrapolation accuracy compared to baseline methods, yielding the lowest MAE and highest PSNR (\Cref{fig:ext_bars}), even when evaluated in a zero-shot manner on the unseen OASIS dataset. Note that Seq-VXM cannot carry out future predictions; thus, it is excluded from this comparison. In addition, the Dice score metric reported in~\Cref{tab:adni_dice,fig:ext_dice_box} further confirms, in agreement with the intensity-based metrics, that TimeFlow achieves better anatomical region alignment.
Qualitative extrapolation results are visualized in~\Cref{fig:qual} (Year 8) and~\Cref{fig:qual_ext_left}, illustrating TimeFlow's accuracy by comparing to the ground truth images. Moreover, \Cref{fig:qual_ext_right} demonstrates extrapolation at a large $t$ ($t=2,4,6$) where ground truth is not available. In these cases, NODER produces unrealistic and anatomically implausible deformations, resulting in significant errors in future image prediction. Similarly, VXM/\textsubscript{diff}, which applies linear scaling to displacement/velocity fields, generates unsmooth deformations (indicated by high NDV) and dissolves the brain, causing the anatomical details to fade. In contrast, TimeFlow/\textsubscript{diff} consistently generates smooth and coherent extrapolated deformation fields, quantitatively supported by the lowest MAE and minimal NDV values in~\Cref{tab:reg_results}.

\paragraph{\textbf{Given future scan, assess the brain aging}} As emphasized in~\Cref{sec:intro}, the extrapolation capability of TimeFlow enables annotation-free disease progression analysis (i.e., requiring no anatomical segmentation or surface reconstructions for quantification). We compare the brain aging rates across diagnostic groups, i.e., Controls (CN), \gls{MCI}, and Dementia, by predicting the optimal time parameter ($t_\textrm{ext}$) corresponding to the observed future MRI scans. 
Specifically, MRI triplets are sampled from validation and test subsets, where the intervals between the first and second visits match closely the intervals between the second and third visits. The diagnosis at the third visit labels each triplet. Using the first two scans as input, TimeFlow extrapolates forward to the third, future scan. By comparing predictions against actual observed future, we compute the extrapolated time parameter $t_\textrm{ext}$, and divide it by the normalized chronological intervals (in years) to yield a brain aging rate. An aging rate above 1 indicates accelerated pathological changes, as expected in patients with neurodegenerative conditions. Indeed, the boxplot of \Cref{fig:progress} (Left) clearly shows that Dementia patients exhibit the highest progression rate, followed by MCI subjects, while the control group maintains progression rates around 1. The analysis validates that our annotation-free brain aging rate estimate is able to separate disease groups.

\begin{figure*}[t]
    \centering
    \includegraphics[width=\linewidth]{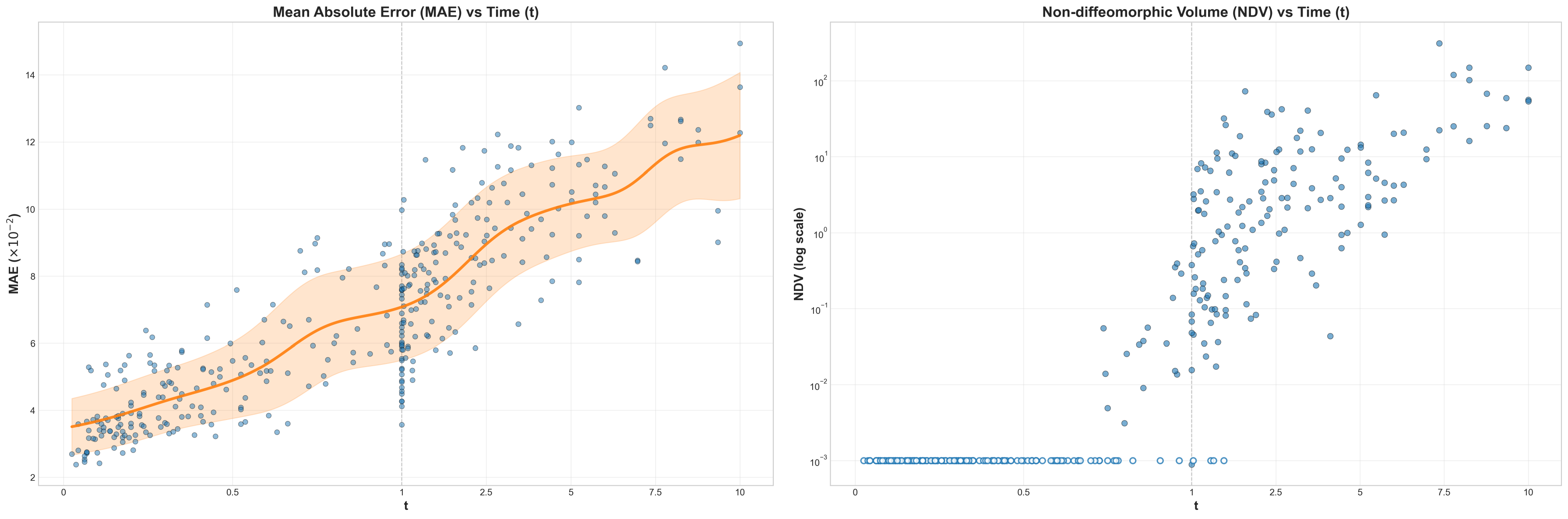}
    \caption{\textbf{Stability analysis across interpolation and extrapolation regimes.} \textbf{Left:} Mean Absolute Error (MAE, in units of $\times 10^{-2}$) of TimeFlow\textsubscript{diff} as a function of time $t$. The solid line shows the Gaussian kernel-smoothed mean, and the shaded region represents $\pm 1$ standard deviation across subjects. \textbf{Right:} Non-diffeomorphic Volume (NDV, logarithmic scale) of TimeFlow\textsubscript{diff} as a function of time $t$. Hollow dots indicate NDV $= 0$ (perfectly diffeomorphic). 
    The vertical dashed line at $t=1$ separates interpolation ($0 < t < 1$) from extrapolation ($t > 1$). The x-axis uses a piecewise linear scale: the left segment spans $t \in [0, 1]$ and the right segment spans $t \in [1, 10]$.}
    \label{fig:mae_ndv_t}
    \vspace{-0.15cm}
\end{figure*}
\begin{figure*}[h]
    \centering
    \includegraphics[width=\linewidth]{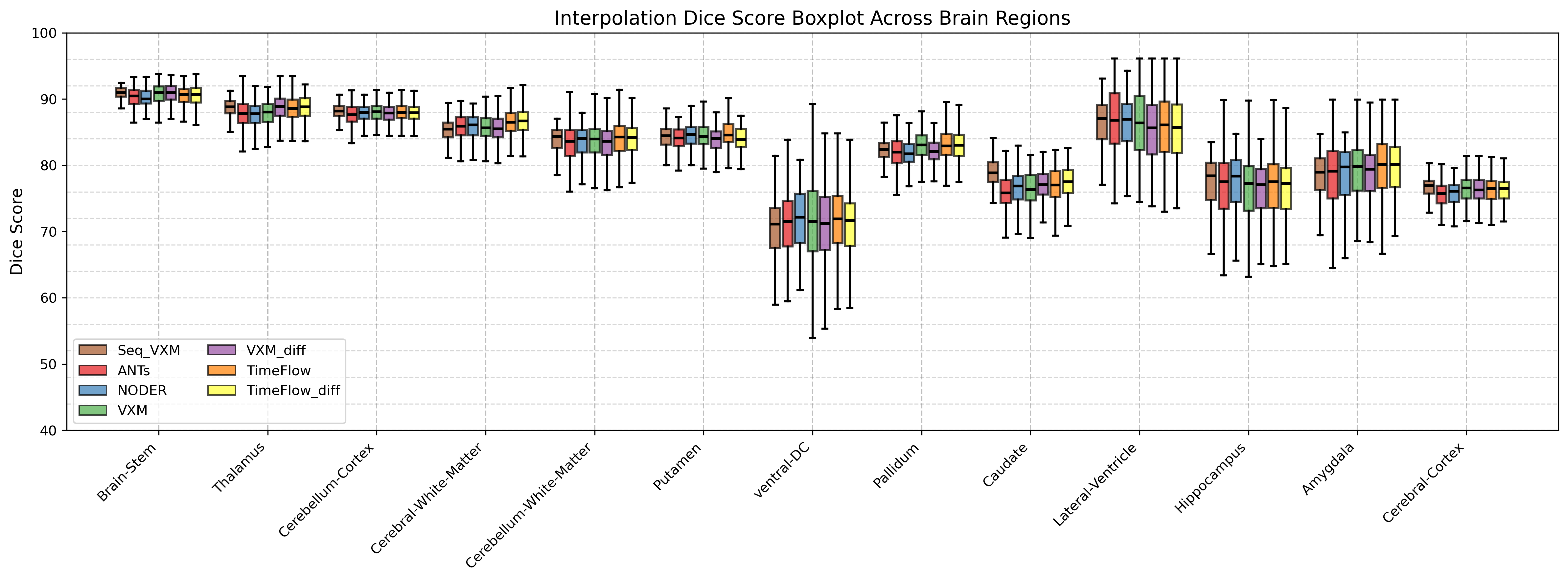}
    \caption{Region-wise Dice score comparison for \textbf{interpolation} on ADNI dataset. Each boxplot shows the distribution of Dice scores across subjects for major anatomical structures. All methods achieve high Dice scores ($>80\%$), with TimeFlow variants performing comparably to the upper-bound Seq-VXM.}
    \label{fig:inter_dice_box}
\end{figure*}

\begin{table*}[ht]
\centering
\setlength{\tabcolsep}{2.5pt}
\caption{Ablation study assessing the impact of inter-/extra-polation image similarity and flow consistency constraints. Temporal smoothness is measured by the Dirichlet energy of the temporal derivative of the time-dependent deformation fields $\|\partial \phi_t /\partial t\|$. Registration performance in terms of MAE (in scale $\times 10^{-2}$), PSNR, SD$\log$J (in scale $\times 10^{-2}$) and non-diffeomorphic volume (NDV) in the foreground brain area are reported. The similarity and flow consistency losses of interpolation and extrapolation are added subsequently for both \textit{TimeFlow} (upper) and \textit{TimeFlow\textsubscript{diff}} (bottom).}
\begin{tabular}{ccccccccccccc}
    \toprule
    \multicolumn{2}{c}{$\mathcal{L}_\textrm{inter}$}
    & \multicolumn{2}{c}{$\mathcal{L}_\textrm{ext}$}
    &
    & \multicolumn{4}{c}{Extrapolation}
    & \multicolumn{4}{c}{Interpolation}
    \\
    \cmidrule(lr){1-2} \cmidrule(lr){3-4}
    \cmidrule(lr){6-9} \cmidrule(lr){10-13}
      sim & flow & sim & flow
      & $\|\partial \phi_t /\partial t\|\downarrow$
       &  MAE $\downarrow$ & PSNR $\uparrow$ & SD$\log$J $\downarrow$ & NDV $\downarrow$
       &  MAE $\downarrow$ & PSNR $\uparrow$ & SD$\log$J $\downarrow$ & NDV $\downarrow$
       \\
    \midrule
    \checkmark & & & & 1.58 $\pm$ 0.17 & 16.7$\pm$7.2  &  13.9$\pm$4.1  &  $>10^3$  &  $>10^7$  &  5.3$\pm$1.0  &  23.3$\pm$1.7  &  4.7$\pm$1.3  &  $>10^4$  \\
    \checkmark & \checkmark & & & 0.82 $\pm$ 0.18 & 9.8$\pm$2.2  &  17.6$\pm$2.1  &  9.7$\pm$4.8  &  $>10^4$  &  5.5$\pm$1.0  &  23.0$\pm$1.7  &  1.8$\pm$0.1  &  0.3$\pm$0.7  \\
    \checkmark & \checkmark & \checkmark & & 0.62 $\pm$ 0.12 & 8.7$\pm$1.7  &  18.7$\pm$1.8  &  5.3$\pm$1.4  &  $>10^3$  &  5.5$\pm$1.0  &  23.0$\pm$1.7  &  1.6$\pm$0.1  &  0.1$\pm$0.4  \\
  \rowcolor{lightgray}  \checkmark & \checkmark & \checkmark & \checkmark & 0.60 $\pm$ 0.11 & 8.6$\pm$1.6  &  18.8$\pm$1.7  &  5.0$\pm$1.0  &  $10^3\pm 10^3$  &  5.5$\pm$1.0  &  23.0$\pm$1.7  &  1.6$\pm$0.1  &  0.1$\pm$0.4  \\
    \midrule
    \checkmark & & & & 3.25 $\pm$ 0.25 & 27.5$\pm$16.2  &  11.4$\pm$6.2  &  $>10^2$  &  $>10^6$  &  5.5$\pm$1.1  &  22.8$\pm$1.9  &  6.9$\pm$2.2  &  $>10^2$  \\
    \checkmark & \checkmark & & & 0.82 $\pm$ 0.12 & 10.4$\pm$2.9  &  17.2$\pm$2.7  &  8.4$\pm$2.7  &  $>10^3$  &  5.3$\pm$1.0  &  23.4$\pm$1.8  &  2.5$\pm$0.2  &  0.3$\pm$0.7  \\
    \checkmark & \checkmark & \checkmark & & 0.55 $\pm$ 0.09 & 8.8$\pm$1.8  &  18.6$\pm$1.9  &  6.1$\pm$1.1  &  18.8$\pm$39.4  &  5.3$\pm$1.0  &  23.3$\pm$1.8  &  2.3$\pm$0.2  &  0.0$\pm$0.1  \\
  \rowcolor{lightgray}  \checkmark & \checkmark & \checkmark & \checkmark & 0.56 $\pm$ 0.09 & 8.8$\pm$1.8  &  18.7$\pm$1.9  &  6.0$\pm$1.0  &  6.1$\pm$11.2  &  5.3$\pm$1.0  &  23.3$\pm$1.8  &  2.2$\pm$0.2  &  0.0$\pm$0.0  \\
    \bottomrule
\end{tabular}
\label{tab:ablation}
\end{table*}

\subsection{Retrospective studies: Matching the deformation trajectory}
We further evaluate TimeFlow's interpolation capability, assessing its effectiveness in modeling temporally continuous and anatomically plausible deformation trajectories connecting observed MRI visits. In this setting, TimeFlow uses only the baseline visit $I_0$ and the final visit $I_L$ as inputs, predicting deformation fields $\phi_{t_k}$ for intermediate images $I_k$, where $0<k<L$.
Quantitative comparisons with baseline methods for interpolation are presented in~\Cref{tab:reg_results}. As discussed in baselines, Seq-VXM, performing explicit pairwise registrations for every available timepoint, represents a performance upper bound. Despite using only the baseline and last scan as inputs, TimeFlow outperforms NODER on the ADNI dataset, which is also aware of the entire sequence and parametrized by Neural\gls{ODE}. TimeFlow achieves comparable performance to NODER in zero-shot evaluation on the OASIS dataset, where NODER performs subject-specific optimization on the full image sequence. Additionally, TimeFlow\textsubscript{diff} achieves the least non-diffeomorphic volume without extra spatial smoothness regularization. Region-wise segmentation-based evaluation (\Cref{tab:adni_dice,fig:inter_dice_box}) validates the superior performance of TimeFlow and TimeFlow\textsubscript{diff}. Qualitative visualizations in~\Cref{fig:qual} (Year 1.5) illustrate the accuracy of TimeFlow's interpolations, substantially reducing alignment errors compared to initial misalignments.
Similar to our above prospective analysis, we also perform retrospective, group-wise brain progression analysis using the inferred temporal parameter $t$ of intermediate visits as a quantitative index of brain progression state marker. We extract trajectory segments with the same diagnoses from each subject, and then compute a linear relationship between the estimated brain state ($t$) and chronological age (years since baseline of the segment). The median linear lines of the groups (CN, MCI, Dementia) in~\Cref{fig:progress} (Right) clearly illustrate distinct progression patterns, with Dementia exhibiting the fastest structural changes, MCI intermediate, and CN the slowest.

\subsection{Clinical Validation, Interpretability and Robustness}
\paragraph{\textbf{Correlation with Cognitive Decline}}
To further validate the clinical relevance of TimeFlow's learned trajectories, we assess the association between the model's estimated brain state ($t$) and cognitive performance using Repeated Measures Correlation (rmcorr). This analysis reveals a significant negative correlation ($r=-0.46$, 95\% CI: $[-0.51, -0.40]$) between the inferred brain state ($t$) and Mini-Mental State Examination (MMSE) scores, indicating that greater pathological progression is associated with declining cognitive function. Importantly, this correlation emerges from trajectories constructed purely from anatomical features of the input image pairs, without any supervision from cognitive scores or diagnostic labels during training. This finding demonstrates that TimeFlow's unsupervised deformation modeling effectively captures clinically meaningful, subject-specific disease progression patterns.

\paragraph{\textbf{Distinction from diagnostic classification}}
Unlike supervised diagnostic classification models that predict discrete disease labels, TimeFlow operates as an unsupervised and disease-agnostic framework designed to capture continuous spatiotemporal structural changes. The model receives no explicit disease labels during training. Instead, it learns to model anatomical trajectories from image pairs alone. The strong alignment between TimeFlow's derived progression metrics and diagnostic groups (\Cref{fig:progress}) emerges naturally from the learned deformation fields, confirming that the framework effectively quantifies pathological deviations without disease-specific priors. This unsupervised nature enables TimeFlow to provide granular, interpretable progression quantification that complements categorical diagnoses. Moreover, by generating predicted future brain images, TimeFlow offers clinicians direct visual intuition: "if the subject continues progressing at the current rate, what will their brain look like in, e.g., 2 years?" This capability bridges quantitative biomarkers with interpretable visual assessment, supporting more informed clinical decision-making.

\paragraph{\textbf{Stability of extrapolation}} 
To assess TimeFlow's robustness across both interpolation and long-range extrapolation, we analyze registration accuracy (MAE) and deformation regularity (NDV) as functions of time $t$ in~\Cref{fig:mae_ndv_t}. For interpolation ($0 < t < 1$), TimeFlow achieves consistently low MAE (below $8 \times 10^{-2}$) and diffeomorphic deformations (NDV $= 0$). For extrapolation ($t > 1$), MAE increases monotonically with prediction distance, reflecting the inherent uncertainty in long-term forecasting. However, for near-to-mid-term predictions ($1 < t \leq 5$), the error remains stable below $10 \times 10^{-2}$, comparable to the mean extrapolation MAE reported in~\Cref{tab:reg_results}. Crucially, NDV remains negligible ($<0.01\%$ of total brain volume) for $t \leq 5$, indicating minimal folding artifacts. These results confirm that TimeFlow generates anatomically plausible deformations and maintains stability for clinically relevant prediction horizons.

\subsection{Ablation Study}
As described in~\Cref{sec:method} and supported by both quantitative and qualitative results, TimeFlow achieves temporally continuous and spatially smooth deformation fields without requiring explicit smoothness regularization. To assess the effectiveness of our proposed inter-/extra-polation consistency constraints, we perform detailed ablation studies. To evaluate the temporal smoothness and continuity of TimeFlow, we compute the derivative $\partial \phi_t / \partial t$ using the finite difference method by sampling $t$ at intervals of $\Delta t=0.25$. Similar to the commonly used diffusion regularizer in learning-based registration methods, which computes the Dirichlet energy over spatial gradients, we quantify temporal smoothness as $\|\partial \phi_t / \partial t\|$. Specifically, we compute the norm of the temporal displacement derivatives within the brain regions and then take the mean over all brain voxels as the final metric. As shown in~\cref{tab:ablation}, incorporating interpolation consistency loss significantly improves both temporal and spatial smoothness ($\|\partial \phi_t / \partial t\|$ and NDV drastically reduce) without sacrificing interpolation accuracy. Furthermore, the extrapolation similarity and flow consistency losses enable TimeFlow to generate plausible extrapolated deformation fields for future image estimation. These ablation results demonstrate that TimeFlow inherently learns temporally continuous and spatially smooth deformation fields, achieving an optimal balance between registration accuracy and smoothness without explicit smoothness penalties.

\subsection{Runtime analysis} 
We compare computational efficiency across all methods. Optimization-based methods exhibit substantially longer runtimes: SyN requires approximately 5 minutes per image pair on CPU, while NODER performs subject-specific optimization requiring 1-2 minutes per subject on GPU. In contrast, learning-based methods achieve efficient amortized inference. TimeFlow and TimeFlow\textsubscript{diff} complete registration in 5-7 seconds per subject, VXM and VXM\textsubscript{diff} in 1-2 seconds, and Seq-VXM in 3-5 seconds (all on GPU). Beyond runtime, memory requirements differ significantly for long sequences: NODER and Seq-VXM require over 16 GB GPU memory for sequences exceeding 5 timepoints, whereas TimeFlow operates efficiently with less than 8 GB.

\section{Discussion}\label{sec:conclu_discuss}
\paragraph{\textbf{Contributions}} 
TimeFlow introduces novel interpolation and extrapolation constraints that ensure temporal consistency without explicit temporal smoothness regularization. 
Combined with the time-conditioning mechanism, TimeFlow enables effective extrapolation beyond the observed time frame. 
While we have used U-Net as the base model, our contributions are generic and can be integrated into other registration backbones. Importantly, TimeFlow has promising application potential in areas such as \textit{in silico} drug effectiveness analysis. By extrapolating brain images to future timepoints at the beginning of a treatment, the predicted outcomes can be compared with actual follow-up scans. Discrepancies between predicted and observed results could then be attributed to treatment effects, offering a novel and efficient approach for evaluating therapeutic efficacy in clinical settings.

\paragraph{\textbf{Distinction from Continuous-time and Generative Frameworks}}
Although NeuralODE-based approaches like NODER~\cite{bai2024noder,wu2024nodeo-seq} share the capability of continuous trajectory modeling, they differ fundamentally from TimeFlow in both computational efficiency and modeling flexibility.
NODER parametrizes a time-varying velocity field and require computationally expensive numerical integration of ordinary differential equations (ODEs). In contrast, TimeFlow learns a direct conditional mapping $\phi_t=f_\theta(I_0, I_L, t)$, bypassing explicit ODE solvers to enable computationally efficient, amortized inference. Furthermore, NODER relies on rigid chronological timestamps from the actual image acquisition times, constraining the trajectory to strictly follow the observed temporal sequence. TimeFlow, however, employs synthetic triplet sampling during training, which provides greater modeling flexibility and robustness to irregular temporal intervals commonly encountered in longitudinal datasets. This design enables TimeFlow to perform reliable inter-/extra-polation even when scan intervals vary substantially.
Regarding generative approaches, while TimeFlow shares architectural similarities with diffusion-based generative models (e.g., U-Net with time embeddings), the objectives differ fundamentally. Diffusion models synthesize image intensities via stochastic denoising processes, generating visually realistic images but operating largely as black boxes with limited interpretability. In contrast, TimeFlow predicts deterministic, biophysically plausible deformation fields that enable direct quantification and interpretation of morphological changes. This distinction makes TimeFlow particularly suitable for clinical applications requiring explainable predictions and precise anatomical correspondence tracking.

\paragraph{\textbf{Bidirectional extrapolation}} While we mainly focus on the forward extrapolation (predicting future brain states) in this study, TimeFlow is inherently bidirectional and allows backward extrapolation as well. Backward extrapolation, i.e., traveling back in time and visualizing the brain at earlier, unobserved ages, opens several avenues: (i) reconstructing a pre-symptomatic baseline for patients before neurodegeneration; (ii) generating counterfactual controls that share a subject's anatomy but simulate a healthy aging trajectory.

\paragraph{\textbf{Generalization and Future Outlook}}
The principles underlying TimeFlow extend well beyond the context of brain aging and neurodegeneration. The framework can generalize to other populations, such as the opposite end of the lifespan: the infant brain, which exhibits rapid structural changes during early development. Moreover, the proposed inter-/extrapolation consistency constraints are inherently modality- and organ-agnostic, opening the possibility for application to other imaging modalities (e.g., T2-weighted, FLAIR) or even other organ systems. For example, by adapting the time embedding to capture the cyclic motion of the heart, TimeFlow could potentially model cardiac dynamics throughout the cardiac cycle. We acknowledge, however, that pure registration has limits: pathologies involving distinct intensity changes or new tissue appearance, such as tumors, inflammatory lesions, or acute stroke, require both geometric modeling and intensity synthesis. In such cases, combining TimeFlow's precise deformation fields with generative models for intensity prediction could provide a more comprehensive approach to disease progression modeling.

\paragraph{\textbf{Limitation and failure cases}} TimeFlow encounters challenges in the following two scenarios: 1. Future image prediction/aging progression analysis with minimal biological aging differences: When the two input scans are temporally close and exhibit no noticeable structural differences, they collapse into a single timepoint, which TimeFlow struggles to predict future images effectively.
2. Prediction for far future: Although TimeFlow is designed to estimate future images for any given time parameter $t \in [0, \infty]$, its performance deteriorates for large $t$ values, e.g., $t$ > 6. This limitation is expected due to limited data covering such long-range time intervals in the ADNI dataset, meaning TimeFlow rarely encountered such scenarios during training. The modeling of such long developments, which are influenced by a plethora of parameters, is inherently challenging.

\section{Conclusion} 
We introduced TimeFlow, a novel framework for longitudinal brain MR registration, specifically designed to address persistent challenges such as limited sequential data, the accuracy-temporal consistency trade-off, and the inability to predict future brain states.
TimeFlow leverages a U-Net-based architecture with temporal condition embeddings to ensure temporally continuous and spatially smooth deformation fields, eliminating the need for explicit smoothness penalties. Unlike existing methods that require full sequential data, TimeFlow learns effectively from only two scans per patient, achieving both high registration accuracy and robust extrapolation to future timepoints. 
Extensive evaluations demonstrate TimeFlow's superior performance over existing baselines in both interpolation and extrapolation tasks, as well as its ability to facilitate clinically meaningful analyses of brain aging and neurodegenerative diseases. By seamlessly bridging retrospective interpolation and prospective extrapolation, TimeFlow enables annotation-free analyses, representing a significant step toward data-driven models for tracking and predicting neuroanatomical changes in neurodegenerative diseases.

{
    \bibliographystyle{IEEEtran}
    \bibliography{main}
}
\end{document}